\documentclass[aps,prl,twocolumn,groupedaddress,showpacs]{revtex4}
\usepackage{graphicx}
\usepackage{amssymb}
\usepackage{color}

\begin{document}


\title{Evidence for Ground- and Excited-State Efimov Trimers in a Three-State Fermi Gas}


\author{J. R. Williams, E. L. Hazlett, J. H. Huckans, R. W. Stites, Y. Zhang and K. M. O'Hara}
\affiliation{Department of Physics, Pennsylvania State University,
University Park,\nolinebreak \,Pennsylvania 16802-6300, USA}


\date{\today}

\begin{abstract}
We observe enhanced three-body recombination in an ultracold three-component $^6$Li Fermi gas with large but unequal scattering lengths attributable to an excited Efimov trimer state near the three-atom scattering threshold.  We find excellent agreement between the measured three-body recombination rate and the recombination rate calculated in the zero-range approximation where the only free parameters are the Efimov parameters $\kappa_*$ and $\eta_*$.  The value of $\kappa_*$ determined by the location of the Efimov resonance we observe at 895~G also predicts the locations of loss features previously observed near 130 and 500~G~\cite{Jochim08,OHara09} suggesting that all three features are associated with universal Efimov trimer states.  We also report on the first realization of a quantum degenerate three-state Fermi gas with approximate SU(3) symmetry.
\end{abstract}


\maketitle

%
%

A landmark result of few-body physics is Efimov's solution to the quantum mechanical three-body problem for low-energy, identical particles with a large $s$-wave scattering length $a$~\cite{Efimov1970,Efimov1979}.  Being independent of the particular character of the two-body interaction, Efimov's predictions are universal.  At resonance ($a \rightarrow \pm \infty$), an infinite series of arbitrarily shallow three-body bound states (Efimov trimers) exist with a geometric spectrum: $E_n = -\left(e^{2\pi/s_0}\right)^{-n}\;\hbar^2 \kappa_*^2/m \; ({\mathrm{for}}\;n = 0,1,2,\ldots)$.
Here, $s_0 \approx 1.00624$ is a universal constant, $m$ is the particle's mass, and the entire spectrum is fixed by a single three-body parameter $\kappa_*$ which is determined by short-range interactions~\cite{Hammer06}.  For large but finite $a$, the spectrum of Efimov trimer states are universal functions of $a$ and $\kappa_*$.  Further, all low-energy scattering observables display a log-periodic dependence on $a$ and the collision energy~\cite{Hammer06}.  These startling predictions ignited a search for Efimov physics in a range of physical systems~\cite{Garrido04,Stoll05}.

The first experimental evidence for Efimov trimers was obtained by observing both a resonant enhancement and interference minimum of three-body recombination in an ultracold gas of bosonic Cs near a Feshbach resonance~\cite{Grimm06}.  The Efimov effect in bosonic systems near an isolated Feshbach resonance is now becoming well understood following the observation of atom-dimer resonances~\cite{Grimm09A}, Efimov resonances between bosons with different masses~\cite{Minardi09} and tests of universal log-periodic scaling~\cite{Modugno09,Khaykovich09}.

\begin{figure}
 \includegraphics[width=3.35in]{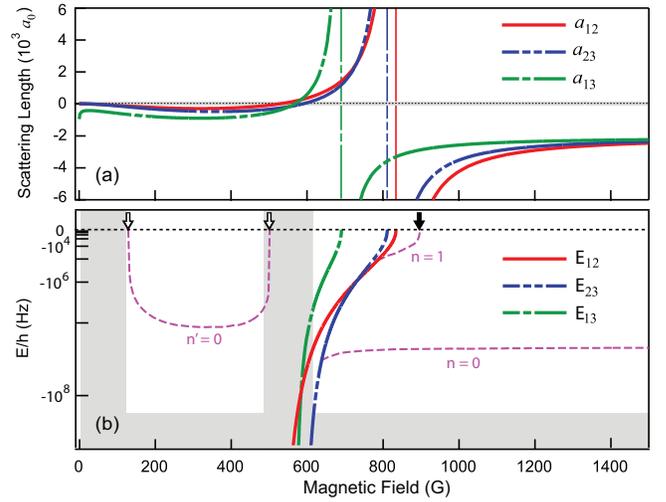} \vspace{-0.12in} 
 \caption{\label{TrimerOverview} (color online) (a) Three overlapping $s$-wave Feshbach resonances in $^6$Li for states $|1\rangle$, $|2\rangle$, and $|3\rangle$~\cite{Julienne05}.  At high field, $a_{ij} \rightarrow -2140 a_0$.  (b) The binding energies ($E_{12}$, $E_{23}$, $E_{13}$) of the universal dimer states associated with the Feshbach resonances.  The dashed lines ($n$ and $n'$) depict the binding energies of the Efimov trimer states (not yet accurately known).  Resonant three-body recombination occurs (arrows) when an Efimov trimer intersects the three-atom scattering threshold.  The grey-shaded areas identify non-universal regions where $E < E_{{\mathrm{vdW}}} = \hbar^2/m \ell_{{\mathrm{vdW}}}^2$ or $|a_{ij}| < 2 \ell_{{\mathrm{vdW}}}$.  (For $^6$Li, $\ell_{{\mathrm{vdW}}} = 62.5\,a_0$.)\vspace{-0.17in}}
\end{figure}

The Efimov effect is also expected to occur for three fermions in distinguishable states ($|1\rangle$, $|2\rangle$, and $|3\rangle$) if at least two of the three pairwise scattering lengths $a_{ij}$ ($a_{12}$, $a_{23}$, and $a_{13}$) are larger than the range of interaction (which, for atoms, is the van der Waals length $\ell_{{\mathrm{vdW}}} = \sqrt[4]{m C_6/\hbar^2}$ )~\cite{Hammer06}.  For a Fermi gas of $^6$Li atoms in the three lowest spin states, investigated here, all three scattering lengths can be resonantly enhanced by three overlapping Feshbach resonances (see Fig~\ref{TrimerOverview} (a))~\cite{Julienne05} allowing for a study of the Efimov effect in a Fermi system which is deeply in the universal regime $|a_{ij}| \gg \ell_{{\mathrm{vdW}}}$.  For this $^6$Li system, the scattering lengths are typically unequal, and any number of them can be positive.  This gives rise to an Efimov spectrum with a rich structure which is only now becoming understood theoretically~\cite{Platter08B,Esry09}.  Entirely new phenomena related to the Efimov effect have been predicted in this case (e.g. interference minima in the rate of atom-dimer exchange reactions)~\cite{Esry09}.

%
%

Beyond few-body physics, superfluid phenomena in degenerate three-component Fermi gases have recently been a subject of intense theoretical interest~\cite{ColorSuperList,Zhai07,FFLOList,Demler07,Zhou08}.  For fermions with $a_{12} \approx a_{23} \approx a_{13} < 0$, there is a competition between trimer formation and several types of Cooper pairing~\cite{ColorSuperList}, breached pair and unbreached pair phases~\cite{FFLOList}, as well as a unique interplay between magnetization and superfluidity~\cite{Demler07,Zhou08}.  Degenerate three-component Fermi gases open a window of opportunity to determine which pairing symmetries and phases are ultimately favored by nature as parameters are tuned.

%
%

In this Letter, we report resonantly enhanced three-body recombination in a three-state Fermi gas with large, negative and unequal scattering lengths that can be explained by an excited Efimov trimer state near the three-atom scattering threshold.  We measure the three-body recombination rate for fields between 842 and 1500~G which is deeply in the universal region at high field (see Fig.~\ref{TrimerOverview}(b)).  The data is well described by a calculation of the three-body recombination rate at threshold in the zero-range approximation for which the only input parameters are the known scattering lengths $a_{12}$, $a_{23}$ and $a_{13}$~\cite{Julienne05} and the complex-valued Efimov parameter $\kappa_* \exp[i \eta_*/s_0]$ which we determine.  Here, $\eta_*$ has been introduced to account for the nonzero width of Efimov states due to decay to deeply bound dimer states.  We find that the value of $\kappa_*$ predicts that Efimov trimers also cross threshold near 130 and 500~G.  This strongly supports the interpretation that the loss features previously observed~\cite{Jochim08,OHara09} are related to the Efimov effect.  Finally, we produce the first degenerate three-state Fermi gas in the SU(3) symmetric regime where $a_{12} \approx a_{23} \approx a_{13}$.

%
%

The first experiments with degenerate three-component Fermi gases investigated their stability against decay via three-body recombination~\cite{Jochim08,OHara09}.  These experiments studied a $^6$Li gas in the three lowest spin states ($|1\rangle$, $|2\rangle$, and $|3\rangle$) and measured the three-body recombination rate as the scattering lengths ($a_{12}$, $a_{23}$ and $a_{13}$) were tuned in a magnetic field.  Enhanced three-body recombination was observed near fields of 130 and 500~G where $|a_{12}|, |a_{23}| \sim \ell_{{\mathrm{vdW}}}$ and $|a_{13}| \sim 10\,\ell_{{\mathrm{vdW}}}$.  Several authors used Efimov's theoretical framework to explain the magnitude and variation of the measured three-body recombination rate between 100 and 500~G despite its questionable applicability near the boundaries of this field range where $|a_{12}|, |a_{23}| \sim \ell_{{\mathrm{vdW}}}$ ~\cite{Platter08B,Ueda08,Wetterich09,Jochim09}.  In this interpretation, the resonant enhancement near 130 and 500~G is due to Efimov-like trimers crossing the three-atom threshold near/in the non-universal regime.

This same three-state $^6$Li mixture can be used to test universal predictions of Efimov physics with large scattering lengths since (1) $|a_{12}|$, $|a_{23}|$ and $|a_{13}| \gg \ell_{{\mathrm{vdW}}}$ for fields above 660~G due to three overlapping Feshbach resonances~\cite{Julienne05} and (2) there is negligible two-body loss~\cite{OHara09}.  The three-body recombination rate in the vicinity of the Feshbach resonances was measured in Ref.~\cite{OHara09}.  However, the authors noted that the highest recombination rates were unitarity limited and therefore could not provide tests of universal predictions~\cite{Esry04}.

%
%

Efimov's theoretical framework makes universal predictions for three-body scattering observables in the threshold regime where the collision energy is the smallest energy in the system.  For identical bosons with large negative scattering lengths, the characteristic energy scale is set by the height of a barrier in the adiabatic three-body potential, $U_{{\mathrm{max}}} = 0.158 \hbar^2/m a^2$~\cite{Esry04}.  In the three-state $^6$Li gas we study, where $a_{12} < a_{23} < a_{13} < 0$, we expect that for fields $B > 875\,{\mathrm{G}}$ ($B > 960\,{\mathrm{G}}$) which corresponds to scattering lengths $|a_{12}| < 12,250 a_0$ ($|a_{12}| < 5,000 a_0)$, a temperature $T \lesssim 30{\mathrm{nK}}$ ($T \lesssim 180 {\mathrm{nK}}$) is required in order to compare our measurements to the calculated recombination rate at threshold.

%
%

In order to reach such low collision energies, we adiabatically release a two-state mixture of $^6$Li atoms prepared in a crossed optical dipole trap (as described in Ref.~\cite{OHara09}) into a larger volume hybrid magnetic/optical trap with smaller oscillation frequencies.  We use two different trapping geometries for the large volume trap to study atoms at $T\simeq 30\,{\mathrm{nK}}$ and $T\simeq180\,{\mathrm{nK}}$ respectively.  For trap A (B), the combination of magnetic and optical (from a 1064 nm dipole trap) forces produce a trap with frequencies $\nu_x = 15(2)\,{\mathrm{Hz}}$, $\nu_y = 0.242 \sqrt{B}\,{\mathrm{Hz}} \pm 1\%$, and $\nu_z = 12(1) \,{\mathrm{Hz}}$ ($\nu_x = 33 \sqrt{1 + 1.4\times 10^{-3} (B - 842\,{\mathrm{G}})}\,{\mathrm{Hz}} \pm 3\%$, $\nu_y = 21 \sqrt{1 + 3.6\times 10^{-3} (B - 842\,{\mathrm{G}})}\,{\mathrm{Hz}}\pm 3\%$, and
$\nu_z = 94(2)\,{\mathrm{Hz}}$)~\cite{TrapFrequencies}.  Here, the dependence of the frequencies on the bias field $B$ (in Gauss) for trap A(B) arises from the nonzero field curvature of a large (small) set of electromagnets used to produce the bias field $B$.

%
%

Starting from an incoherent 50/50 mixture of atoms in states $|1\rangle$ and $|2\rangle$, we create a three-state mixture with equal population in states $|1\rangle$, $|2\rangle$ and $|3\rangle$ by simultaneously applying radio-frequency (RF) magnetic fields which drive $|1\rangle - |2\rangle$ and $|2\rangle - |3\rangle$ transitions at the field $B$.  Each of the RF fields has a resonant Rabi frequency $\Omega/2 \pi \sim 10\,{\mathrm{kHz}}$, is broadened to a width of 10 kHz and applied for a variable time (chosen to be at least twice the observed decoherence time) between $10$ and $100$ ms depending on the bias field.

\begin{figure}
 \includegraphics[width=3.35in]{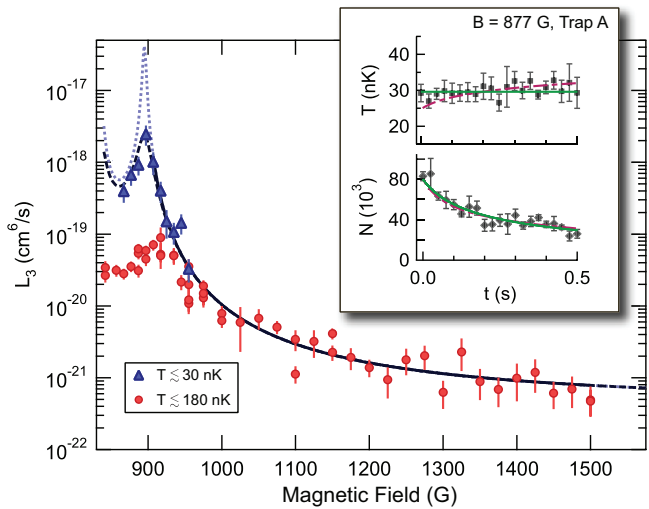}\vspace{-0.12in}
 \caption{\label{ThreeBodyLossRate}  (color online) The measured three-body recombination rate constant $L_3$.  Resonant three-body recombination near 900~G is caused by an excited Efimov trimer crossing the three-atom scattering threshold (see text).  The inset shows typical data for $N(t)$ and $T(t)$ which are fit to extract $L_3$.  In the inset the solid (dashed) line shows the fit to a model which does not include (includes) anti-evaporation.\vspace{-0.17in}}
\end{figure}

%
%
We have confirmed (for the fields studied here) that the decay of each of the possible two-state mixtures of the three states is consistent with one-body loss due to background gas collisions.  One-body loss for these experiments gives a $1/e$ lifetime $1/\Gamma = 2.8\,{\mathrm{s}}$.  In the three-state mixture, three-body recombination involving one atom from each spin state can occur in addition to one-body loss.   In this case, the number of atoms in each of the equally populated spin states $N(t)$ evolves according to $\dot{N}/N = -\Gamma - L_3 \langle n^2 \rangle$ where $L_3$ is the recombination rate constant and $\langle n^2 \rangle$ is the average value of the squared density per spin state.  For a thermal distribution at temperature $T$ in a harmonic trap, $N(t)$ evolves according to
\begin{eqnarray}
\label{NumEvol}
\frac{1}{N} \frac{d N}{d t} & = & - \Gamma - \frac{L_3}{\sqrt{27}} \left(\frac{2 \pi m \bar{\nu}^2}{k_B T}\right)^3 N^2,
\end{eqnarray}
where $\bar{\nu} = \left(\nu_x \nu_y \nu_z \right)^{1/3}$.  Empirically, the temperature of the gas remains approximately constant.  This is consistent with the fact that the energetic atom and molecule produced in a recombination event have a mean free path much larger than the sample size and exit the cloud without depositing energy.  Our measurements are consistent with a model that includes a rise in temperature due to ``anti-evaporation''~\cite{OHara09} (see inset to Fig.~\ref{ThreeBodyLossRate}).  However, lower values of $\chi^2$ are obtained if the temperature is simply assumed to remain constant.

To determine the recombination rate constant $L_3$, we measure for each field $B$ the number and temperature of the trapped atoms as a function of time by {\emph{in situ}} absorption imaging.  We show an example of data recorded at a particular field value in the inset to Fig.~\ref{ThreeBodyLossRate}.  The number evolution at each field is fit with an analytic solution to Eqn.~\ref{NumEvol} using $L_3$, the initial number ($N_0$)  and temperature ($T$) as free parameters.  The recombination rate constant for fields between 834 and 1500~G is shown in Fig.~\ref{ThreeBodyLossRate} for data sets taken with traps A and B.  The triangles (circles) correspond to trap A(B) and give $L_3$ for a cloud at a temperature $\lesssim 30\,{\mathrm{nK}}$ ($\lesssim 180\,{\mathrm{nK}}$) and an initial peak density per spin state $n_0 \simeq 5 \times 10^{9}\,{\mathrm{atoms/cm}}^3$ ($n_0 \simeq 5 \times 10^{10}\,{\mathrm{atoms/cm}}^3$).  The error bars give the statistical error from the fit and the uncertainty in the trap frequencies added in quadrature.  A systematic uncertainty of $\pm60\%$, which arises from our uncertainty in the absolute atom number ($\pm 30\%$), is not included in these error bars.  As we vary the field, the scattering lengths are tuned by broad Feshbach resonances at 690, 811, and 834~G and the observed value of $L_3$ changes by several orders of magnitude.  In addition to a smooth variation of $L_3$, resonant three-body recombination is observed near 900~G in both data sets.

In order to compare our data to zero-temperature calculations of the recombination rate, we require that the temperature be in the threshold regime and that the data is unaffected by the unitarity limit~\cite{Esry04}. As argued above, the threshold regime should be obtained for the 30~nK (180~nK) gas for fields above 875~G(960~G).  For a gas at a low but nonzero temperature $T$, the recombination rate constant for three distinguishable fermions with equal mass is unitarity limited to a maximum value $L_{3 {\mathrm{max}}} =  \alpha/(k_B T)^2$ where $\alpha = \sqrt{108} \pi^2 \hbar^5/m^3$~\cite{Esry04,Burke99}.  For a thermal gas with $T = 30\,{\mathrm{nK}}$ ($T = 180\,{\mathrm{nK}}$), $L_{3 {\mathrm{max}}} = 8 \times 10^{-18}\,{\mathrm{cm}}^6/{\mathrm{s}}$ ($L_{3 {\mathrm{max}}} = 2 \times 10^{-19}\,{\mathrm{cm}}^6/{\mathrm{s}}$).  The measured values of $L_3$ are $\lesssim (1/10) L_{3{\mathrm{max}}}$ for the 30~nK (180~nK) data for fields above 907~G (975~G).  These subsets of the data should be in excellent agreement with zero-temperature calculations of the recombination rate.

%
%
Recently, Braaten and co-workers calculated the three-body recombination rate constant in the zero-range approximation for three fermions in distinguishable spin states at threshold~\cite{Platter08B}.  By numerically solving a generalization of the ${\mathrm{Skorniakov-}}{\mathrm{Ter-}}{\mathrm{Martirosian}}$ (STM) equation, they calculate the recombination rate constant for the case  $a_{12},\,a_{23},\,a_{13} < 0$.  The only inputs to this calculation are $a_{12}$, $a_{23}$ and $a_{13}$ (known for $^6$Li~\cite{Julienne05}) and the two three-body parameters, $\kappa_*$ and $\eta_*$, which respectively fix the spectrum and linewidth of the Efimov trimer states.  For equal scattering lengths, the STM equation reduces to that for identical bosons and, in this case, an analytic expression for $L_3$ is known~\cite{Platter08B}
\begin{eqnarray}
\label{L3forEqualA}
L_3 = \frac{16 \pi^2 C \sinh (2 \eta_*)}{\sin^2\left[ D \left| a \right| \kappa_*\right] + \sinh^2\eta_*} \frac{\hbar a^4}{m} \; (\mathrm{for}\,a_{ij} = a < 0).
\end{eqnarray}
Here the constants $C = 29.62(1)$ and $D = 0.6642(2)$ have been determined from numerical calculations~\cite{Platter08B}.  The analytic expression above exhibits resonant enhancement at values $a = a_n^- = -(e^{\pi/s_0})^n (D \kappa_*)^{-1}$ for integer $n$ which occur when Efimov trimer states cross the three-atom threshold.  For $a_{12} \neq a_{23} \neq a_{13}$, as in our experiment, the STM equations must be solved numerically~\cite{DaekyoungKangCode} to accurately determine $L_3$.

Since $|a_{12}|, |a_{23}|, |a_{13}| \gg \ell_{vdW}$ and $a_{12}, a_{23}, a_{13} < 0$ for fields above 834~G, the data shown in Fig.~\ref{ThreeBodyLossRate} should be well described by the zero-range calculation of $L_3$ in Ref.~\cite{Platter08B} and will allow the first determination of the three-body parameters for this three-component $^6$Li gas in the high-field regime. The best fit (solid line) to subsets of the 30~nK data ($B \geq 907\,{\mathrm{G}}$) and the 180~nK data ($B \geq 975\,{\mathrm{G}}$) (as described above) is shown in Fig.~\ref{ThreeBodyLossRate}.  The only free parameters in this fit are $\kappa_*$ and $\eta_*$.  From the fit we obtain $\kappa_* = 6.9(2)\times10^{-3} a_0^{-1}$ and $\eta_* = 0.016^{+0.006}_{-0.010}$ where the uncertainties -- combined statistical and systematic -- indicate one standard deviation.  The zero-temperature recombination rate for these three-body parameters calculated for fields between 840 and 1550~G is shown as the dotted line in Fig.~\ref{ThreeBodyLossRate}.  In this model, the resonant peak in the recombination rate occurs at 895~G where an Efimov trimer crosses the three-atom scattering threshold.  The measured three-body recombination rate for the 30\,nK data also peaks near 895~G, though at a significantly smaller magnitude due to the unitarity limit.  The dashed line shows a ``unitarized'' recombination rate~\cite{Suno04} given by $\left(1/L_3 + 1/L_{3{\mathrm{sat}}}\right)^{-1}$ to account for the observed saturation of the recombination rate in the thermal gas to a value $L_{3{\mathrm{sat}}} \approx L_{3{\mathrm{max}}}/3$.

We can identify the Efimov trimer that crosses threshold at 895~G with the first excited state of the Efimov spectrum.  The binding energy of this Efimov trimer, in the limit of infinite scattering lengths, is $E_1 = e^{-2 \pi/s_0} \hbar^2 \kappa_*^2/m = h \times 55(3)\,{\mathrm{kHz}}$.  In this limit, the ground state Efimov trimer has a binding energy a factor of $e^{2 \pi/s_0} \approx 515$ larger, $E_0 = h \times 28(2)\,{\mathrm{MHz}}$.
This is the lowest state with a binding energy smaller than the van der Waals energy scale $E_{\mathrm{vdW}} = \hbar^2/m \ell_{\mathrm{vdW}}^2$.
If it were the case that $a_{12} = a_{23} = a_{13}=a < 0$, the  first excited trimer would cross threshold at $a = a_1^- = -5.0(1) \times 10^3\,a_0$ and the adjacent trimer states would cross threshold at values of $a$ either larger or smaller by a factor of $e^{\pi/s_0} \approx 22.7$.  In reality, an Efimov trimer of the $^6$Li system crosses threshold at 895~G where $a_{12} = -8584\,a_0$, $a_{23} = -5702\,a_0$ and $a_{13} = -2893\,a_0$.  We identify this trimer as the first excited Efimov trimer since $a_1^-$ falls within this range of $a_{ij}$ values.  In the high-field region ($B > 608\,{\mathrm{G}}$), the ground state Efimov trimer asymptotes to a binding energy $E_0(a_{ij} = -2140\,a_0) = h \times 24(2)\,{\mathrm{MHz}}$ with a width $\gamma = (2 \pi) \, 1.5_{-1}^{+0.6}\,{\mathrm{MHz}}$ (corresponding to a $100^{+180}_{-24}$~ns lifetime) as calculated from expressions given in Ref.~\cite{Hammer06}.

%
%

One aspect of Efimov physics that can now be tested with ultracold atom experiments is the extent to which the universal results can be applied when scattering lengths are tuned across poles or across nodes.  For the $^6$Li gas studied here, there are two universal regions ($122~{\mathrm{G}} < B < 485{\mathrm{G}}$ and $B > 608\,{\mathrm{G}}$) where the zero-range approximation should be applicable, separated by a non-universal region where the scattering lengths become smaller than $2\,\ell_{{\mathrm{vdW}}}$.  We find that the value $\kappa_* = 6.9(2)\times10^{-3} a_0^{-1}$, determined above for the high-field region, is close to the value $\kappa_* = 6.56\times 10^{-3}\,a_0^{-1}$ previously determined from the position of the narrow loss feature in the low-field region near 130~G~\cite{Platter08B}.  Indeed, the recombination rate calculated in the zero-range approximation using $\kappa_* = 6.9(2)\times10^{-3} a_0^{-1}$ predicts Efimov resonances at $125(3)$ and $499(2)\,{\mathrm{G}}$ in reasonable agreement with observations~\cite{Jochim08,OHara09}.  These resonances occur when the ground state Efimov trimer crosses threshold.  The value of $\eta_*$ is nearly 1 order of magnitude larger in the low field region~\cite{Platter08B} than in the high-field regime.  The significant change in $\eta_*$ is likely due to the dramatic difference in the binding energy of the deep dimer states for the two regions~\cite{Jochim09}.

%
%
Finally, we note that despite the large three-body recombination rate that occurs in this system, we can produce quantum degenerate three-component Fermi gases in the high-field regime where the three scattering lengths are large, negative and approximately equal.  Starting from a degenerate two-state mixture of $^6$Li atoms at $B = 1500\,{\mathrm{G}}$ in trap B, we prepare a three-state mixture by applying RF pulses as described above. After equilibration, the three-state mixture has $N = 6(2)\times 10^4$ atoms per spin state, a temperature $T = 50(10)\,{\mathrm{nK}}$, a Fermi temperature $T_F = 180(20)\,{\mathrm{nK}}$ and a degeneracy temperature $T/T_F = 0.28(6)$.  The difference in mean field energies is more than one order of magnitude smaller than any other energy scale in the system making it useful for future investigations of three-component Fermi gases with SU(3) symmetry~\cite{ColorSuperList,Zhai07,Demler07,Zhou08,FFLOList}.

%
%

We are indebted to E. Braaten and D. Kang for providing the computer code to numerically calculate $L_3$.  We gratefully acknowledge fruitful conversations with E. Braaten, B. Esry, J. D'Incao, and D. Kang  regarding this work.
This material is based upon work supported by the AFOSR (Award FA9550-08-1-0069), the ARO (Award W911NF-06-1-0398) and
the NSF (PHY 07-01443).

During preparation of this manuscript we became aware that enhanced recombination in a $^6$Li gas near 900~G was also observed by S. Jochim's group~\cite{Jochim09}.


\begin{thebibliography}{28}
\bibitem[{Tra()}]{TrapFrequencies}
\bibinfo{journal}{We determine the frequencies by observing dipole oscillations or, for $\nu_z = 94~{\mathrm{Hz}}$, by parametric excitation.}
\end{thebibliography}
\end{document}